\documentclass[aps,prd,twocolumn,superscriptaddress]{revtex4-1}
\usepackage[]{mathtools}
\usepackage{bm}
\usepackage{graphicx}
\usepackage{subfigure} 
\usepackage{float}
\usepackage[colorlinks,linkcolor=blue,anchorcolor=blue,citecolor=blue]{hyperref} 
\usepackage{verbatim}
\usepackage{booktabs}
\usepackage{multirow}
\usepackage{color}

\newcommand{\lwen}[1]{\textcolor{black}{#1}}

\begin{document}


\title{Semianalytical Approach for Sky Localization of Gravitational Waves}

\author{Qian Hu}
\email{hq2017@mail.ustc.edu.cn}

\author{Cong Zhou}
\affiliation{CAS Key Laboratory for Research in Galaxies and Cosmology, Department of Astronomy, University of Science and Technology of China, Hefei 230026, China}
\affiliation{School of Astronomy and Space Sciences, University of Science and Technology of China, Hefei, 230026, China}

\author{Jhao-Hong Peng}
\affiliation{Department of Physics, National Taiwan Normal University, 88, Section 4, Ting-Chou Rd, Taipei 116, Taiwan}
\author{Linqing Wen}
\email{linqing.wen@uwa.edu.au}
\author{Qi Chu}
\author{Manoj Kovalam}
\affiliation{Australian Research Council Centre of Excellence for Gravitational Wave Discovery (OzGrav), The University of Western Australia, 35 Stirling Highway, Crawley, Western Australia 6009, Australia}
\affiliation{Department of Physics, The University of Western Australia, 35 Stirling Highway, Crawley, Western Australia 6009, Australia}

\date{\today}

\begin{abstract}
Rapid sky localization of gravitational wave sources is crucial to enable prompt electromagnetic follow-ups. In this article, we present a novel semianalytical approach for sky localization of gravitational waves from compact binary coalescences. We use the Bayesian framework with an analytical approximation to the prior distributions for a given astrophysical model.  We derive a semianalytical solution to the posterior distribution of source directions. This method only requires one-fold numerical integral that marginalizes over the merger time, compared to the five-fold numerical integration otherwise needed in the Bayesian localization method. The performance of the method is demonstrated using a set of  \textcolor{black}{binary neutron stars (BNS)} injections on Gaussian noise using LIGO-Virgo's design and O2 sensitivity. \textcolor{black}{\lwen{We find the m}edian of 90\% confidence area in O2 sensitivity \lwen{to be} $\mathcal{O}(10^2)~\mathrm{deg^2}$, comparable to that of the existing LIGO-Virgo online localization method \texttt{Bayestar} and parameter estimation toolkit \texttt{LALInference}. In the end, we apply this method to localize the BNS event GW170817 and \lwen{find} the 50\% (90\%) confidence region of $11~\mathrm{deg^2}$ ($50~\mathrm{deg^2}$). The detected optical counterpart of GW170817 \lwen{resides within} our 50\% confidence area.}

\end{abstract}

\maketitle

\section{\label{sec1} Introduction}

Since the first detection of gravitational wave (GW) signals by the Laser Interferometer Gravitational-Wave Observatory (LIGO) in 2015~\cite{Abbott:2016blz, Harry:2010zz}, there have been over 50 GW detections~\cite{gwtc1, o3a,LIGOScientific:2021usb,zackay2019detecting,PhysRevD.101.083030,PhysRevD.100.023011,PhysRevD.100.023007,Nitz_2019,Nitz_2020,nitz20213ogc,Magee_2019,virgoO3,PhysRevLett.123.231107,PhysRevD.102.062003} from \textcolor{black}{Advanced LIGO and Advanced Virgo}~\cite{2015, Acernese_2014,PhysRevD.93.112004,PhysRevLett.116.131103}. An exceptional discovery was made in 2017 when a GW detection of a binary neutron star (BNS) coalescence (GW170817)~\cite{TheLIGOScientific:2017qsa} was linked to a gamma-ray burst (GRB) observation and other electromagnetic (EM) follow-up observations~\cite{Abbott2017a}, answering a series of questions including the long-standing conjecture of the progenitor for GRB~\cite{Savchenko:2017ffs}. In the latest 3rd LIGO and Virgo observational run (O3), a new initiative was setup to send GW alerts in real time in the hope to capture such GW-EM coincident event~\footnote{\url{https://gcn.gsfc.nasa.gov/gcn3/24045.gcn3}}. Speedy GW early warning, with reasonably accurate localization,  is considered to be the key to capture more of such events~\cite{chichi2016,Magee_2021,Cannon_2012,Sachdev_2020,Nitz_2020fore}. 

The localization of GW sources is a crucial step for joint GW-EM observations. The conventional parameter estimation method \texttt{LALInference}~\cite{Veitch2015a} uses a general Bayesian framework and uses the MCMC, or the Nested Sampling techniques to sample the entire parameter space and marginalizes nuisance parameters to obtain the source directions. This usually takes hours to days to finish, \textcolor{black}{ which is not practical for rapid source localization of GWs. }

\textcolor{black}{\lwen{F}ast parameter estimation methods are proposed to tackle this problem~\cite{PhysRevD.92.023002,Cornish2021a,PhysRevD.103.043011,PhysRevD.101.022003,Singer2016}, including \texttt{Bayestar} algorithm~\cite{Singer2016} which is currently used to} provide fast online localization of GW sources after a GW signal is detected and uploaded to the LVC database. A quintuple numerical integral is used to yield the posterior distribution of the GW source directions by marginalizing over source parameters of individual masses, spins of individual component, and five extrinsic parameters (distance, binary inclination, polarization angle, coalescence phase and merger time).

In this work, we provide a new semianalytical solution to the posterior distribution of the GW source directions. Specifically, we start from the Bayesian theorem and marginalize over 5 extrinsic parameters semianalytically.  We test our algorithm on injected GW signals with simulated Gaussian noise for the LIGO-Virgo's 2nd science run (O2) and for the future design sensitivity of LIGO and Virgo. This analytical result is expected to help reduce the computational cost and the resulting latencies of source localization, and with the potential to be implemented in a coherent online search, e.g., in the SPIIR pipeline~\cite{Hooper:2012aa,Hooper:2012zz,chichi,Liu:2012vw}.

\section{\label{sec2} Bayesian method for localization}
Bayesian method is a conventional statistical method for parameter estimation and it is widely used in the GW field~\cite{Thrane2019}. According to Bayes' theorem, giving the prior probability distribution of a set of parameters $\boldsymbol{\vartheta}$, one can obtain the updated distribution, the posterior distribution, given the observed data set $\mathbf{d(t)}$. Mathematically, the theorem is expressed as:
\begin{equation}
    \label{bayestheo}
    \underbrace{p(\boldsymbol{\vartheta} \mid \mathbf{d}(t))}_{\text {Posterior }}
    =
    \frac{\overbrace{p(\mathbf{d}(t) \mid \boldsymbol{\vartheta})}^{\text {Likelihood }}
    \overbrace{p(\boldsymbol{\vartheta} )}^{\text {Prior }}} {\underbrace{p(\mathbf{d}(t) )}_{\text {Evidence }}},
\end{equation}
The evidence can be considered constant in our estimation, thus posterior is proportional to the product of the likelihood and the prior distribution.

\subsection{Signal model}
To solve the Bayes posterior, we first model our likelihood function. We assume the additive Gaussian noise $n$ for each detector, the detector data $d$ when a GW signal $h$ is present can be expressed as:
\begin{equation}
    d^{(i)}(t) = h^{(i)}(t;\boldsymbol{\vartheta}) + n^{(i)}(t),
\end{equation} 
where superscript $(i)$ denotes the i-th detector, $\boldsymbol{\vartheta}$ is the parameter set. The binary coalescence GW signal parameters can be divided into the \textit{intrinsic} parameters and the \textit{extrinsic} parameters as shown in Tab.~\ref{tab1}. As the focus of this work is on the rapid localization of the binary neutron star events, the templates used are not taking into account the eccentricity, the tidal deformability or the spin parameters. 
\begin{table}
\begin{tabular}{l|c|l}
    \hline 
    \multirow{4}{*}{\text{ Intrinsic parameters}} & $m_{1}$ & Mass of first body \\
    & $m_{2}$   & Mass of second body \\
    & $\mathbf{S}_{1}$ & Spin of first body \\
    & $\mathbf{S}_{2}$ & Spin of second body \\
    \hline  & $\alpha$  & Right ascension angle\\
    & $\delta$ & Declination angle \\
    & $r$ & Distance \\
    Extrinsic parameters & $\iota$ & Inclination angle \\
    & $t_{c}$ & Arrival time at detector \\
    & $\psi$ & Polarization angle \\
    & $\phi_{c}$ & Coalescence phase \\
    \hline
\end{tabular}
\caption{\label{tab1} Intrinsic and extrinsic parameters of the GWs from compact binary coalescence systems. }
\end{table}

In more detail, a detector response to a GW signal is a linear combination of the two GW polarizations plus ($h_+$) and cross ($h_{\times}$). A GW signal present in a detector can be written as:
\begin{equation}
    \label{hfhf}
    h(t)^{(i)} = F^{(i)}_+(\alpha, \delta, \psi, t_c)h_+(t) + F^{(i)}_{\times}(\alpha, \delta, \psi, t_c)h_{\times}(t),
\end{equation}
where $F^{(i)}_{+,\times}(\alpha, \delta, \psi, t_c)$ are \textit{detector beam-pattern functions}~\cite{GWbook1} for i-th detector which rely on the right ascension $\alpha$, the declination $\delta$, the polarization angle $\psi$ and the coalescence time $t_c$. 
To separate $\psi$ and $\alpha, \delta$, we define \textit{amplitude modulation functions} $G^{(i)}_{+,\times}$~\cite{chichi}
\begin{equation}
    G^{(i)}_{+,\times}(\alpha, \delta, t_c) = F^{(i)}_{+,\times}(\alpha, \delta, \psi=0, t_c),
\end{equation}
Expressions of $G^{(i)}_{+,\times}$ can be found in Eq.~(1.53-1.54) in Ref.~\cite{chichi}. Although they are dependent on time, they are treated as constants during the GW events as the time durations of GW signals for ground-based detectors are short compared to the self-rotation period of the Earth~\cite{findchirp}. 

The two polarizations can be expressed by the extrinsic amplitude evolution $a(t)$ and the phase evolution $\phi(t)$ with the inclination angle $\iota$, and the coalescence phase $\phi_c$~\cite{chichi}:
\begin{equation}
    \label{PNwaveform}
    \begin{array}{l}
    h_{+}(t)=\frac{1}{2}a(t)(1+\cos ^{2} \iota)  \cos \left(\phi(t)+\phi_{c}\right) \\
    h_{\times}(t)=a(t) \cos \iota \sin \left(\phi(t)+\phi_{c}\right).
    \end{array}
\end{equation} If we define the two quadrature functions:
\begin{equation}
    \begin{array}{l}
    h_{c}(t)=a(t, r=1\mathrm{Mpc}) \cos (\phi(t)), \\
    h_{s}(t)=a(t, r=1\mathrm{Mpc}) \sin (\phi(t)),
    \end{array}
\end{equation}
the GW signal in a given detector can be expressed as:
\begin{equation}
    \label{eq:rearranged}
    h^{(i)} = (G_+^{(i)}, G_{\times}^{(i)}) \mathbf{A_c} h_c + (G_+^{(i)}, G_{\times}^{(i)}) \mathbf{A_s} h_s,
\end{equation}
where 
\begin{widetext}
\begin{equation}
    \label{Amatrix}
    \begin{aligned}
    \mathbf{A} &= 
    \begin{pmatrix}
        \mathbf{A_c} && \mathbf{A_s}
    \end{pmatrix}
    =
    \begin{pmatrix}
    A_{11} && A_{12}\\
    A_{21} && A_{22}
    \end{pmatrix}\\
    &=\frac{1\mathrm{Mpc}}{r} \begin{pmatrix} \cos2 \psi & \sin 2\psi \\ -\sin2 \psi & \cos2\psi \end{pmatrix} \begin{pmatrix} \frac{1+\cos^{2}\iota}{2} &  \\  & \cos\iota \end{pmatrix} \begin{pmatrix} \cos\phi_{c} & \sin\phi_{c} \\ -\sin\phi_{c} & \cos\phi_{c}. \end{pmatrix}
    \end{aligned}
\end{equation}
\end{widetext}

The likelihood function for one detector can be written as~\cite{Finn_1992}:
\begin{equation}
    \label{LR}
    \begin{aligned}
        p(d^{(i)}(t) \mid \boldsymbol{\vartheta}) &\propto e^{-(d^{(i)}-h^{(i)}|d^{(i)}-h^{(i)})/2} \\
        &\propto e^{(d^{(i)}|h)-\frac{1}{2}(h^{(i)}|h^{(i)})}.
    \end{aligned}
\end{equation}
The inner product between two time series is defined as:
\begin{equation}
    (a \mid b)=4 \mathcal{R} \int_{0}^{\infty} \frac{\tilde{a}(f) \tilde{b}^{*}(f)}{S_{n}(f)} d f,
\end{equation}
where \textit{tilde} denotes frequency domain and \textit{star} means complex  conjugate. $S_{n}(f)$ is one-sided power spectral density (PSD) that evaluates the noise in a detector, which is defined by 
\begin{equation}
    \left\langle\tilde{n}(f) \tilde{n}^{*}\left(f^{\prime}\right)\right\rangle=\frac{1}{2} S_{n}(|f|) \delta\left(f-f^{\prime}\right).
\end{equation}
Here $\left\langle \dots \right\rangle$ is ensemble average.

With a network of detectors, we define the inner product for two arrays or two matrices of time series as:
\begin{equation}
    \mathbf{C} = (\mathbf{D} |\mathbf{B} ) \Rightarrow C_{j k}=\sum_{p=1}^{n}\left(D_{j p} \mid B_{p k}\right),
\end{equation}
where $\mathbf{D}$ is an $m\times n$ matrix of time series, $\mathbf{B}$  is an $n\times l$ matrix and the result $\mathbf{C}$ is an $m\times l$ matrix.

The likelihood for a detector network can be expressed as:
\begin{equation}
    \label{eq:LR1}
        p(\mathbf{d} \mid \boldsymbol{\vartheta})
        \propto e^{(\mathbf{d^T}|\mathbf{h})-\frac{1}{2}(\mathbf{h^T}|\mathbf{h})}
\end{equation}
where 
\begin{equation}
    \mathbf{d}=
\begin{pmatrix}
d^{(1)}(t+\tau^{(1)})  \\
d^{(2)}(t+\tau^{(2)}) \\
\vdots \\
d^{(N)}(t+\tau^{(N)}) \\
\end{pmatrix},
\quad
\mathbf{h}=
\begin{pmatrix}
h^{(1)}(t+\tau^{(1)})  \\
h^{(2)}(t+\tau^{(2)}) \\
\vdots \\
h^{(N)}(t+\tau^{(N)}) \\
\end{pmatrix},
\end{equation}
where $\tau^{(i)}$ is to take into account different arrival times of the signal. Extending Eq.~\ref{eq:LR1} with Eq.~\ref{eq:rearranged} and split the matrix $\mathbf{A}$ into $\mathbf{A_c}$ and $\mathbf{A_s}$, we have the network likelihood function:
\begin{equation}
    \label{eq:LR2}
        p(\mathbf{d} \mid \boldsymbol{\vartheta})
        \propto \prod_{x=\{c, s\}} e^{ \left(\mathbf{d}^{\mathbf{T}} \mid \mathbf{G} \mathbf{A}_{\mathbf{x}} h_{x}\right)-\frac{1}{2}\left(\mathbf{A}_{\mathbf{x}}^{\mathbf{T}} \mathbf{G}^{\mathbf{T}} h_{x} \mid \mathbf{G} \mathbf{A}_{\mathbf{x}} h_{x}\right)} \\
\end{equation}  
$h_c$ and $h_s$ are in quadrature so the cross-correlation is 0 and we define $sigma$ as the inner product:
\begin{equation}
    \begin{aligned}
     (h_c | h_s) &= 0 \\
     \sigma^{(i)} \equiv \sqrt{(h_c | h_c)} \mid_{r=1\mathrm{Mpc}} &= \sqrt{(h_s | h_s)} \mid_{r=1\mathrm{Mpc}}
    \end{aligned}
\end{equation}
then the likelihood becomes
\begin{equation}
    \label{eq:LR3}      
    p(\mathbf{d} \mid \boldsymbol{\vartheta})
        \propto \prod_{x=\{c, s\}} e^{\left(\mathbf{d}^{\mathbf{T}} \mid \mathbf{H}_{\mathbf{x}}\right) \mathbf{G}_{\mathbf{\sigma}} \mathbf{A}_{\mathbf{x}}-\frac{1}{2} \mathbf{A}_{\mathbf{x}}^{\mathbf{T}} \mathbf{G}_{\mathbf{\sigma}}^{\mathbf{T}} \mathbf{G}_{\mathbf{\sigma}} \mathbf{A}_{\mathbf{x}}} ~.
\end{equation}
where $\mathbf{G_{\sigma}}$ is  
\begin{equation}
    \mathbf{G_{\sigma}} = 
\begin{pmatrix}
G_{+}^{(1)}\sigma^{(1)} & G_{\times}^{(1)}\sigma^{(1)} \\
G_{+}^{(2)}\sigma^{(2)} & G_{\times}^{(2)}\sigma^{(2)} \\
\vdots & \vdots \\
G_{+}^{(N)}\sigma^{(N)} & G_{\times}^{(N)}\sigma^{(N)}
\end{pmatrix}.
\end{equation}
and $\mathbf{H_{c,s}}$ is the normalized signal given as:
\begin{equation}
    \mathbf{H_{c,s}} = \mathrm{diag}\left(\frac{h_{c,s}}{\sigma^{(1)}},  \frac{h_{c,s}}{\sigma^{(2)}}, \dots, \frac{h_{c,s}}{\sigma^{(N)}} \right),
\end{equation}

where \textit{diag} denotes diagonal matrix. $\mathbf{H_{c,s}}$ is also used to compute the matched filtering signal-to-noise ratio (SNR), a widely-used statistic in GW detection:
\begin{equation}
    \label{mfoutput}
    \boldsymbol{\rho} = ( \mathbf{H_{c}} \mid \mathbf{d} ) + i( \mathbf{H_{s}} \mid \mathbf{d} ).
\end{equation}
The i-th element of $\boldsymbol{\rho}$ is the SNR time series of the i-th detector. It is a complex time series and we use its modulus as the SNR. The network SNR is defined as $\sqrt{\boldsymbol{\rho^T}\boldsymbol{\rho^*}}$.

\section{\label{sec3} Parameter choice and Prior setting}
\subsection{Parameter Choice}
Considering the errors of the intrinsic parameters, including the binary masses, are semi-independent from errors in sky localization~\cite{Singer2016}, we therefore set component masses as the values determined by matched filtering. This is reasonable for the purpose of achieving low-latency online localization of GW sources. 

We only need to consider the extrinsic parameters for our sky direction estimation.
By rearranging extrinsic parameters $(\iota, \phi_c, r, \psi)$, the likelihood can be rewritten (Eq.~\ref{eq:LR3}) as a function of $\mathbf{A}$ (Eq.~\ref{Amatrix}) with these extrinsic parameters included implicitly. We therefore replace them with $\mathbf{A}$. Moreover, we divide the sky into  equal areas when calculating posterior probability, thus it is more convenient to use $\sin \delta$ than $\delta$. To sum up, we adopt the following parameter transformation:
\begin{equation}
    \label{trans}
    (t_c, \alpha, \delta, \iota, \phi_c, r, \psi) \rightarrow (t_c, \alpha, \sin \delta, A_{11}, A_{21}, A_{12}, A_{22}).
\end{equation}
and from Eq.~\ref{Amatrix}, we have
\begin{equation}
\begin{aligned}
    \label{aijexplicit}
    A_{11} &= \frac{1}{r}\left(\frac{1+\cos^2\iota}{2} \cos 2\psi \cos \phi_c - \cos \iota \sin 2\psi \sin\phi_c\right), \\
    A_{21} &= -\frac{1}{r}\left(\frac{1+\cos^2\iota}{2} \sin 2\psi \cos \phi_c + \cos \iota \cos 2\psi \sin\phi_c\right), \\
    A_{12} &= \frac{1}{r}\left(\frac{1+\cos^2\iota}{2} \cos 2\psi \sin \phi_c + \cos \iota \sin 2\psi \cos\phi_c\right), \\
    A_{22} &= -\frac{1}{r}\left(\frac{1+\cos^2\iota}{2} \sin 2\psi \sin \phi_c - \cos \iota \sin 2\psi \cos\phi_c\right).
    \end{aligned}
\end{equation}

\subsection{Prior setting by Monte-Carlo Simulation}
The prior distribution in Eq.~\ref{bayestheo} should be the joint distribution of the 7 extrinsic parameters $p(\alpha, \sin \delta, t_c,\mathbf{A})$. However, we can assume $\alpha$, $\sin \delta$, $t_c$ and $\mathbf{A}$ are independent to each other, i.e.
\begin{equation}
    \label{prior-inde}
    p(\alpha, \sin \delta, t_c,\mathbf{A}) = p(\alpha)p(\sin \delta)p(t_c)p(\mathbf{A}).
\end{equation}

We employ general prior distributions on $\alpha$, $\sin \delta$, $t_c$ since we have no information for them in advance of source localization. We assume the GW source is isotropic in the sky and the coalescence time is uniformly distributed in $\pm 10$ms around the trigger time. Namely,
\begin{equation}
    \begin{aligned}
    p(\alpha)&\propto 1,\\
    p(\sin \delta)& \propto 1,\\
    p(t_c)&\propto 1,
    \end{aligned}
\end{equation}
which means they can be treated as constants in the posterior probability density.

\begin{figure*}
    \includegraphics[width=1\textwidth]{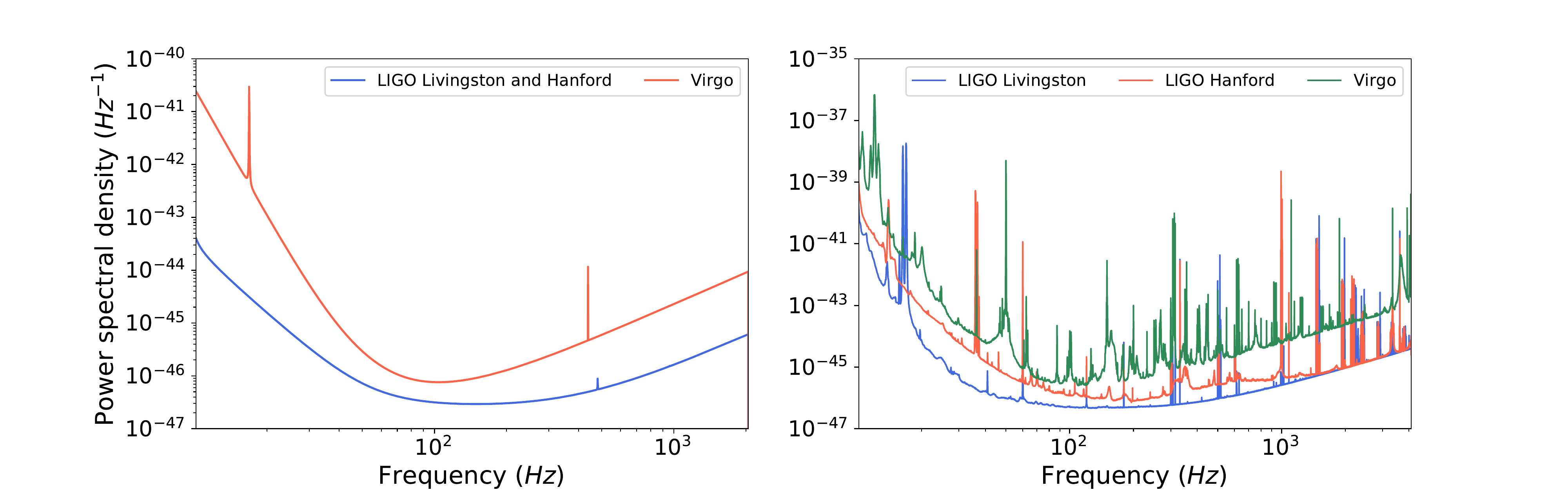}
    \caption{\label{o2psd} Left panel: PSDs used at design sensitivity from LALSuite~\cite{lalsuite}. The PSD is generated with low frequency cutoff flow corresponding to the aLIGO 2016-2017 high-sensitivity scenario in LIGO-P1200087~\cite{designPSD}. We adopt the same design sensitivity for the LIGO Livingston and LIGO Hanford observatories. Right panel: PSDs of LIGO Livingston, LIGO Hanford and Virgo in the second observation run (O2) \textcolor{black}{generated \lwen{using} data \lwen{from GW Open Science Center~\cite{GWOSC}.} 
    }}
\end{figure*}

To investigate properties of $\mathbf{A}$ from the original four extrinsic parameters $(\iota, \phi_c, r, \psi)$, we simulate 50000 BNS events and inject them to simulated random Gaussian noise for 3 detectors, \textcolor{black}{LIGO Hanford, LIGO Livingston and Virgo (HLV)}. Two different kinds of sensitivities have been used, one is the design sensitivity and the other is the second observation run (O2) sensitivity as shown in Fig.~\ref{o2psd}.

Waveforms are generated using the TaylorT4~\cite{TaylorT4} approximant with zero spins. Component masses are uniformly distributed between 1.3 and 1.5~$M_{\odot}$. Sky positions and BNS orientations are drawn from isotropic distribution. Distance is drawn from a uniform distribution in volume with the maximum distance at 200~Mpc. Polarization angle and coalescence phase are uniformly sampled between [$0,\pi$] and [$0,2\pi$], respectively. We ignore cosmological effects as the redshift for LIGO-Virgo Detectors is only 0.044 at 200~Mpc for the standard cosmology. 

\begin{figure*}
    \includegraphics[width=1\textwidth]{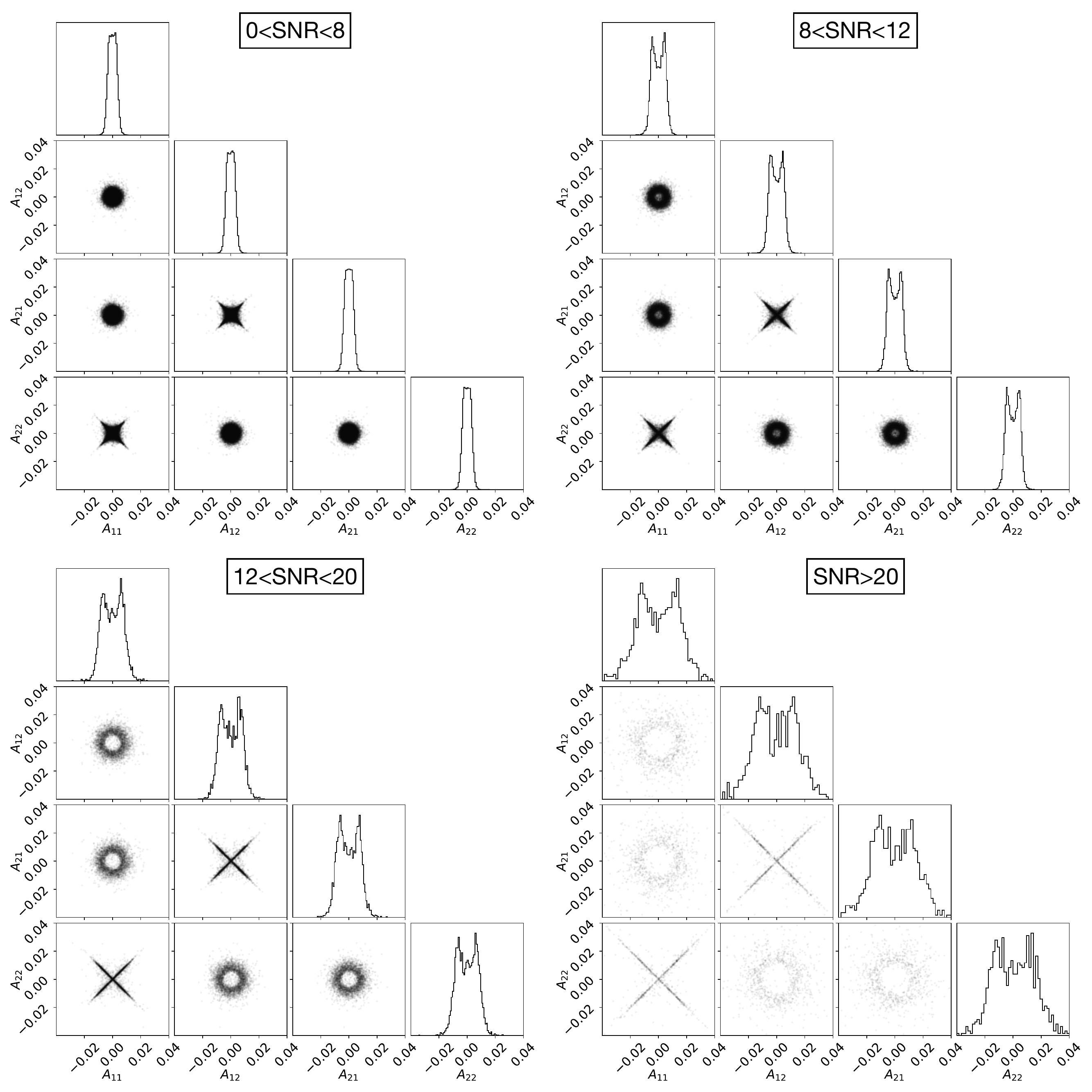}
    \caption{\label{corralationfig}  Corner plot of elements of $\mathbf{A}$ generated from 50000 simulations in \textcolor{black}{Gaussian noise colored to O2 sensitivity}. We categorized $A_{ij}$ samples according to the HLV network SNR and ploted them in 4 subfigures. }
\end{figure*}

We generate the maximum network SNR by matched filtering and calculate $A_{ij}$ ($i,j=1,2$, $A_{ij}$ represents each element of $\mathbf{A}$ ) by Eq.~\ref{aijexplicit} for each simulated GW event. The corner plot of $A_{ij}$ in different network SNR ranges are shown in Fig.~\ref{corralationfig}. Note here we only show $A_{ij}$ from O2 sensitivity, but the same analysis can be employed on other PSDs. According to Fig.~\ref{corralationfig}, we find the following characteristics of the distribution of $A_{ij}$:
\begin{itemize}
    \item Each $A_{ij}$ follows the similar distribution.
    \item When SNR is high ($>8$), $A_{ij}$ follows bimodal distribution, and the location of the peak depends on SNR.
    \item The diagonal and off-diagonal elements, $(A_{11},A_{22})$ and $(A_{21},A_{12})$, are correlated, while other elements are entirely uncorrelated with each other.
\end{itemize}

Based on the first point, we assume all $A_{ij}$s follow the same distribution. Since a GW signal with network SNR $<8$ is usually not considered as a successful detection, we ignore the low-SNR cases and focus on the bimodal distribution of $A_{ij}$. We adopt a symmetric bimodal prior distribution for $A_{ij}$ with a superposition of two Gaussian functions:
\begin{equation}
    \label{prior}
    p(A_{ij}) \propto e^{-\frac{(A_{ij}-\mu)^2}{2\sigma ^2}} + e^{-\frac{(A_{ij}+\mu)^2}{2\sigma ^2}},
\end{equation}
where $\mu$ and $\sigma$ will be derived numerically for a given astrophysical model of the extrinsic parameters for different ranges of by SNR. 

\begin{figure*}
    \includegraphics[width=1\textwidth]{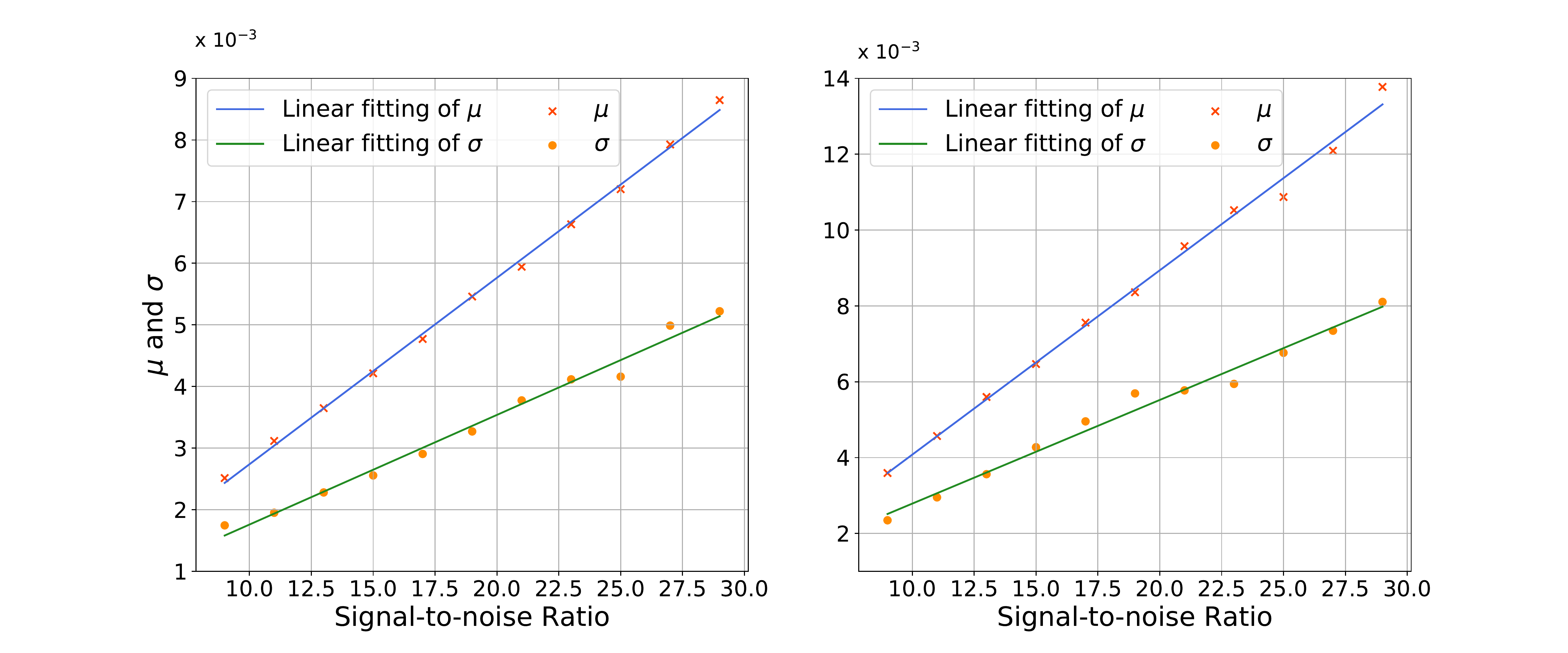}
    \caption{\label{linear_relation_fig} Linear relation between the best-fit $\mu,\sigma$ and SNR. Left panel shows the result in design sensitivity and right panel is for O2. Points of the best-fit values are plotted at the center of their corresponding SNR bin. }
\end{figure*}

For distributions of $A_{ij}$ in each SNR bin with the length of 2, we use the least square method to obtain the best-fit $\mu$ and $\sigma$. Compare the best-fit values in different SNR bins, we find $\mu$ and $\sigma$ have a linear relation with the network SNR, as shown in Fig.~\ref{linear_relation_fig}. For the design sensitivity, we have
\begin{equation}
    \label{emp1}
    \begin{aligned}
        \mu &= 0.0003026~\mathrm{SNR} - 0.0002882,\\
        \sigma &= 0.0001779~\mathrm{SNR} - 0.00001968,
    \end{aligned}
\end{equation}
for O2 sensitivity it becomes
\begin{equation}
    \label{emp2}
    \begin{aligned}
        \mu &= 0.0004860~\mathrm{SNR} - 0.0007827,\\
        \sigma &= 0.0002733~\mathrm{SNR} +0.00005376.
    \end{aligned}
\end{equation}

\begin{figure*}
    \subfigure[Design noise]{\includegraphics[width=0.7\textwidth]{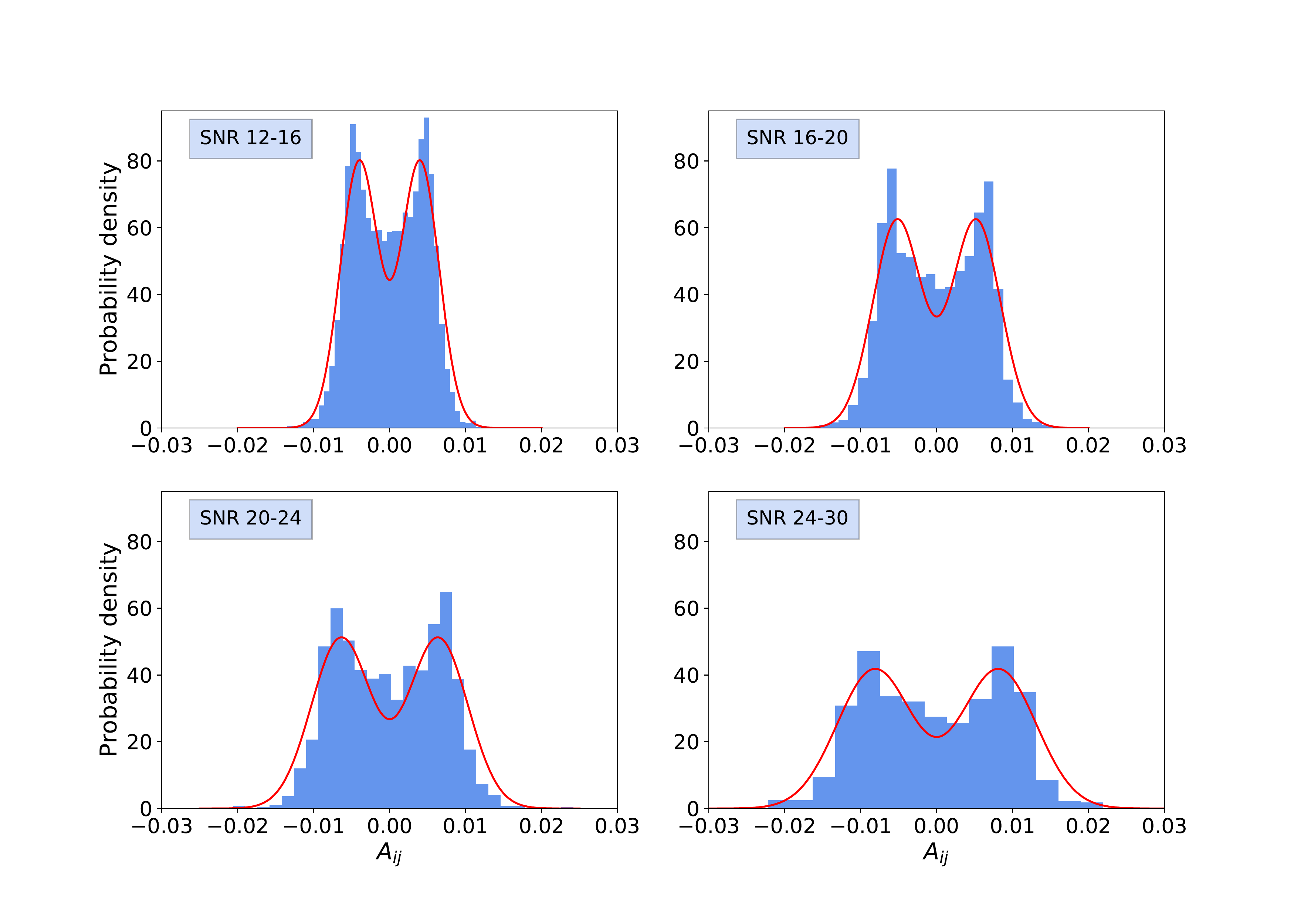}}
    \subfigure[O2 noise]{\includegraphics[width=0.7\textwidth]{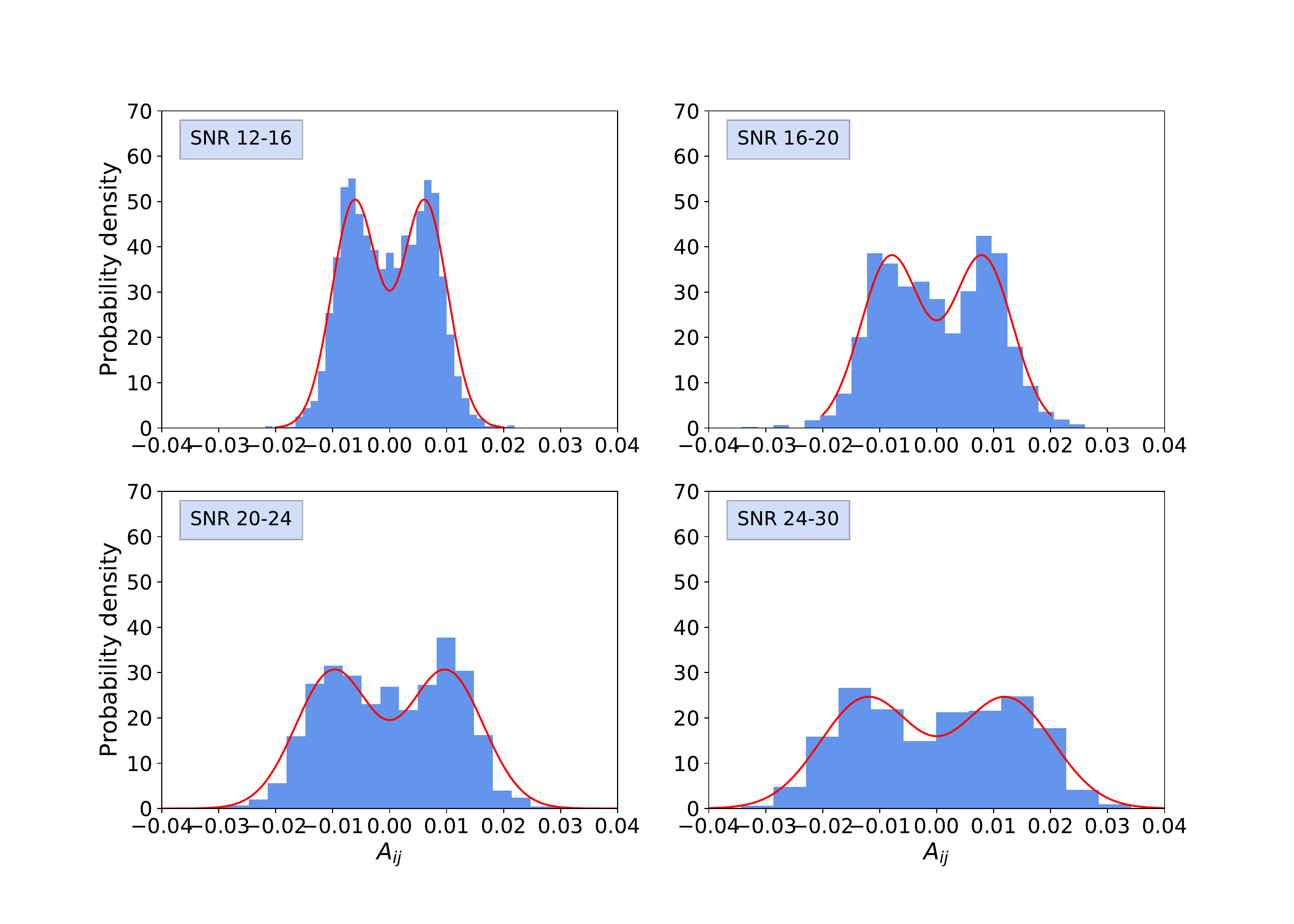}}
    \caption{\label{test_empir_rela} Comparison between $A_{ij}$'s simulated distribution and empirical approximation. Blue bars are $A_{ij}$'s distribution in our simulation; four subplots represent four different SNR ranges. Red lines are empirical distribution with SNRs equal to center value of SNR bins. }
\end{figure*}

Fig.~\ref{test_empir_rela} shows the comparison between the distribution of $A_{ij}$ in different SNR bins and the bimodal prior distribution calculated using Eq.~\ref{emp1} or \ref{emp2} at the central value of the SNR bin. The bimodal distribution with empirical relation is sufficient to reconstruct $A_{ij}$s' distribution. 

The bimodal distribution of $A_{ij}$ originates from the selection effect of high SNR GW events. The amplitude of GWs increases monotonically with $|\cos \iota|$ and $1/r$, therefore, when we select GW events in a higher SNR range, sources with larger $|\cos\iota|$ values and smaller $r$ are more likely to be selected. This also implies that $\cos \iota$ tends to have a symmetric bimodal distribution~\cite{Jaranowski_1998}, which results in the bimodal distribution of $A_{ij}$ (Eq.~\ref{aijexplicit}).  Physically, the bimodality of $\cos \iota$ originates from opposite handedness of the binary orbit, which corresponds to two different inclination angles: $\iota$ and $\pi-\iota$.   On the other hand, small distance $r$ corresponds to larger values of $A_{ij}$, hence the peaks in $A_{ij}$'s distribution move outside. This is why $A_{ij}$ follows bimodal distribution and is dependent on SNR. 

The correlation between $A_{ij}$s can be explained as follows. The correlation is also caused by our selection of high SNR events. As discussed above, for high SNR events we have $|\cos \iota|\rightarrow 1$. According to Eq.~\ref{aijexplicit}, when $\cos \iota \rightarrow 1$, $A_{11} \rightarrow A_{22}, ~A_{12} \rightarrow -A_{21}$; when $\cos \iota \rightarrow -1$, $A_{11} \rightarrow -A_{22}, ~A_{12} \rightarrow A_{21}$. Plus $\cos \iota$ has the same probability to be $1$ or $-1$, as a result, two symmetric crosses are showed in each corner plot of Fig.~\ref{corralationfig}.

As most $A_{ij}$ samples in Fig.~\ref{corralationfig}  are positioned at the diagonal cross, we further assume that $A_{22}$ has half chance to be $A_{11}$ and another half to be $-A_{11}$ for detectable GW events whose SNRs are usually high, i.e.,
\begin{equation}
    \label{condprob1}
    p(A_{22}|A_{11}) = \frac{\delta(A_{22} - A_{11}) + \delta(A_{22}+A_{11})}{2},
\end{equation}
where $\delta()$ is the $\delta$ function: $\delta(0)=1$ and $\delta(x)=0$ otherwise. For the same reason,
\begin{equation}
    \label{condprob2}
    p(A_{12}|A_{21}) = \frac{\delta(A_{21} - A_{12}) + \delta(A_{21}+A_{12})}{2}.
\end{equation}
\textcolor{black}{This approximation is also adopted in previous work on fast GW source localization~\cite{Tsutsui_2021}, in which the authors show the probability distribution of  $\cos \iota$ of detectable GW events 
has strong peaks for \lwen{$\cos \iota = \pm 1$}. 
} Note the correlation only exists in diagonal and off-diagonal elements of $\mathbf{A}$, we have
\begin{equation}
    \label{prior_of_A}
    \begin{aligned}
        p(\mathbf{A}) &= p(A_{11}, A_{12}, A_{21}, A_{22}) \\
        &=p(A_{11}, A_{22})p(A_{21}, A_{12}) \\
        &= p(A_{11})p(A_{22}|A_{11}) p(A_{21})p(A_{12}|A_{21}) ,
    \end{aligned}
\end{equation}
where $p(A_{11})$ and $p(A_{21})$ are given in Eq.~\ref{prior} while $p(A_{22}|A_{11})$ and $p(A_{12}|A_{21})$ are given in Eq.~\ref{condprob1} and~\ref{condprob2}.

\section{\label{sec4} Posterior marginalization}
Rewrite posterior probability here:
\begin{equation}
    p(\boldsymbol{\vartheta} \mid \mathbf{d}) \propto p(\mathbf{d} \mid \boldsymbol{\vartheta}) p(\boldsymbol{\vartheta}) ,
\end{equation}
where likelihood $p(\mathbf{d} \mid \boldsymbol{\vartheta})$ and prior probability $p(\boldsymbol{\vartheta})$ are given by Eq.~\ref{eq:LR3} and Eq.~\ref{prior-inde}, respectively. To obtain the sky localization, we can now marginalize $t$ and elements of $\mathbf{A}$, i.e.,
\begin{equation}
    \label{posterior_int}
    p(\alpha,\delta \mid \mathbf{d}) \propto \int p(\mathbf{d} \mid \boldsymbol{\vartheta}) p(\boldsymbol{\vartheta}) d^4\mathbf{A}dt,
\end{equation}
where $d^4\mathbf{A}= dA_{11}dA_{12}dA_{21}dA_{22}$. Introduce the following Gaussian integral, which gives the integral of Gaussian functions~\cite{gaussianinte}:
\begin{equation}
    \label{wiki}
    \int_{- \infty}^{+\infty} e^{-\frac{1}{2} \mathbf{x}^{\mathbf{T}} \mathbf{A} \mathbf{x}+\mathbf{B}^{\mathbf{T}} \mathbf{x}} d^{2} \mathbf{x}=\sqrt{\frac{(2 \pi)^{2}}{\operatorname{det} \mathbf{A}}} e^{\frac{1}{2} \mathbf{B}^{\mathbf{T}} \mathbf{A}^{-1} \mathbf{B}},
\end{equation}
where $\mathbf{A}$ is $2\times 2$ symmetric and positive definite matrix, $\mathbf{x}$ and $\mathbf{B}$ are $2\times 1$ vectors. $d^{2} \mathbf{x} = dx_1dx_2$. Define 
\begin{equation}
    \label{defm}
    \mathbf{M} = \mathbf{G}_{\sigma}^{\mathrm{T}} \mathbf{G}_{\sigma},
\end{equation}
\begin{equation}
    \label{defj}
    \mathbf{J}_{\mathbf{x}}^{\mathrm{T}}=\left(\mathbf{d}^{\mathrm{T}} \mid \mathbf{H}_{\mathbf{x}}\right) \mathbf{G}_{\sigma},
\end{equation}
likelihood can be converted to the same form as the integrand in Eq.~\ref{wiki}:
\begin{equation}
    \label{LR_re}
    \begin{aligned}
        p(\mathbf{d} \mid \boldsymbol{\vartheta}) &\propto \prod_{x=\{c, s\}} e^{\left(\mathbf{d}^{\mathbf{T}} \mid \mathbf{H}_{\mathbf{x}}\right) \mathbf{G}_{\mathbf{\sigma}} \mathbf{A}_{\mathbf{x}}-\frac{1}{2} \mathbf{A}_{\mathbf{x}}^{\mathbf{T}} \mathbf{G}_{\mathbf{\sigma}}^{\mathbf{T}} \mathbf{G}_{\mathbf{\sigma}} \mathbf{A}_{\mathbf{x}}} \\
        &=\prod_{x=\{c, s\}}e^{-\frac{1}{2} \mathbf{A}_{\mathbf{x}}^{\mathrm{T}} \mathbf{M} \mathbf{A}_{\mathbf{x}}+\mathbf{J}_{\mathbf{x}}^{\mathrm{T}} \mathbf{A}_{\mathbf{x}}} .
    \end{aligned}
\end{equation}

Since both likelihood and prior function are Gaussian function, their product is still a Gaussian functions, which can be integrated by Eq.~\ref{wiki}. Substituting likelihood Eq.~\ref{LR_re} and prior function Eq.~\ref{prior-inde} into posterior probability Eq.~\ref{posterior_int}, and using Gaussian integral Eq.~\ref{wiki}, we get

\begin{equation}
    \label{result}
    \begin{aligned}
    & p\left(\alpha, \delta \mid \mathbf{d}\right) \\
    & \propto \int_{t_{\mathrm{trigger}}-T}^{t_{\mathrm{trigger}}+T} d t_{c} \int d^{4} \mathbf{A} \exp \left\{\sum_{x=\{c, s\}}-\frac{1}{2} \mathbf{A}_{\mathbf{x}}^{\mathrm{T}} \mathbf{M} \mathbf{A}_{\mathbf{x}}+\mathbf{J}_{\mathbf{x}}^{\mathrm{T}} \mathbf{A}_{\mathbf{x}}\right\} \\
    &~ \times p(A_{11}) p(A_{22}|A_{11}) p(A_{21}) p(A_{12}|A_{21})\\
    & \propto \int_{t_{\mathrm{trigger}}-T}^{t_{\mathrm{trigger}}+T} d t_{c} (I_1+I_2+I_3+I_4).
    \end{aligned}
\end{equation}
where $I_{1,2,3,4}$ have analytical expressions and are given in  Appendix~\ref{apdx}. $t_\mathrm{trigger}$ is the trigger time and $T=10$~ms. Eq.~\ref{result} gives the semianalytical posterior probability of the source direction which only needs one-fold numerical integral.

\section{\label{sec5} Performance Test and Case Study}
We apply the proposed localization algorithm (Eq.~\ref{result}) on \textcolor{black}{ \lwen{a new set of randomly generated} GW events 
\lwen{with the same detector} sensitivity described in Sec.~\ref{sec3}. 
} We use the matched filtering technique to search the simulated GW data and set the HLV network SNR (Eq.~\ref{mfoutput}) $\ge$ 12 as the criterion of  successful detection. As a demonstration of its performance on real detector data, we also apply this localization method on the SNR time series output from the SPIIR pipeline~\cite{spiir_paper} of the detected BNS event GW170817. 

\subsection{Confidence and Searched Area}
We define the 90\% and 50\% confidence areas the same way as in \texttt{Bayestar}~\cite{Singer2016}. Using adaptive HEALPix sampling~\cite{Gorski:2005aa,Singer2016}, we first divide the sky into $N_{\mathrm{pix,0}} = 3072$ pixels and calculate the posterior probability of each pixel through Eq.~\ref{result} and we assume the posterior probability is constant within a pixel, thus the probability for each pixel is equal to the calculated posterior probability multiplied by the area. The top $N_{\mathrm{pix}}/4$ most probable pixels are further divided into $N_{\mathrm{pix}}$ daughter pixels, and posterior probability is calculated again for those pixels. We also repeat this process seven times as in \texttt{Bayestar}~\cite{Singer2016}. We then rank the probability of all pixels in descending order and search from the first pixel. The probability of searched pixels is accumulating while the search, until the cumulative sum is equal to the given values of $0.9$ and $0.5$. The area in that credible level is given by the sum of the area of searched pixels. 

We also define a searched area. Searched area is computed by searching from the first pixel defined in the last paragraph until the sky direction of the injection signal is included. Searched area is the smallest of such constructed area that contains the true sky direction  of the source. It measures the accuracy of the localization independently of the precision~\cite{Singer2016}. 

\begin{figure*}
    \includegraphics[width=1\textwidth]{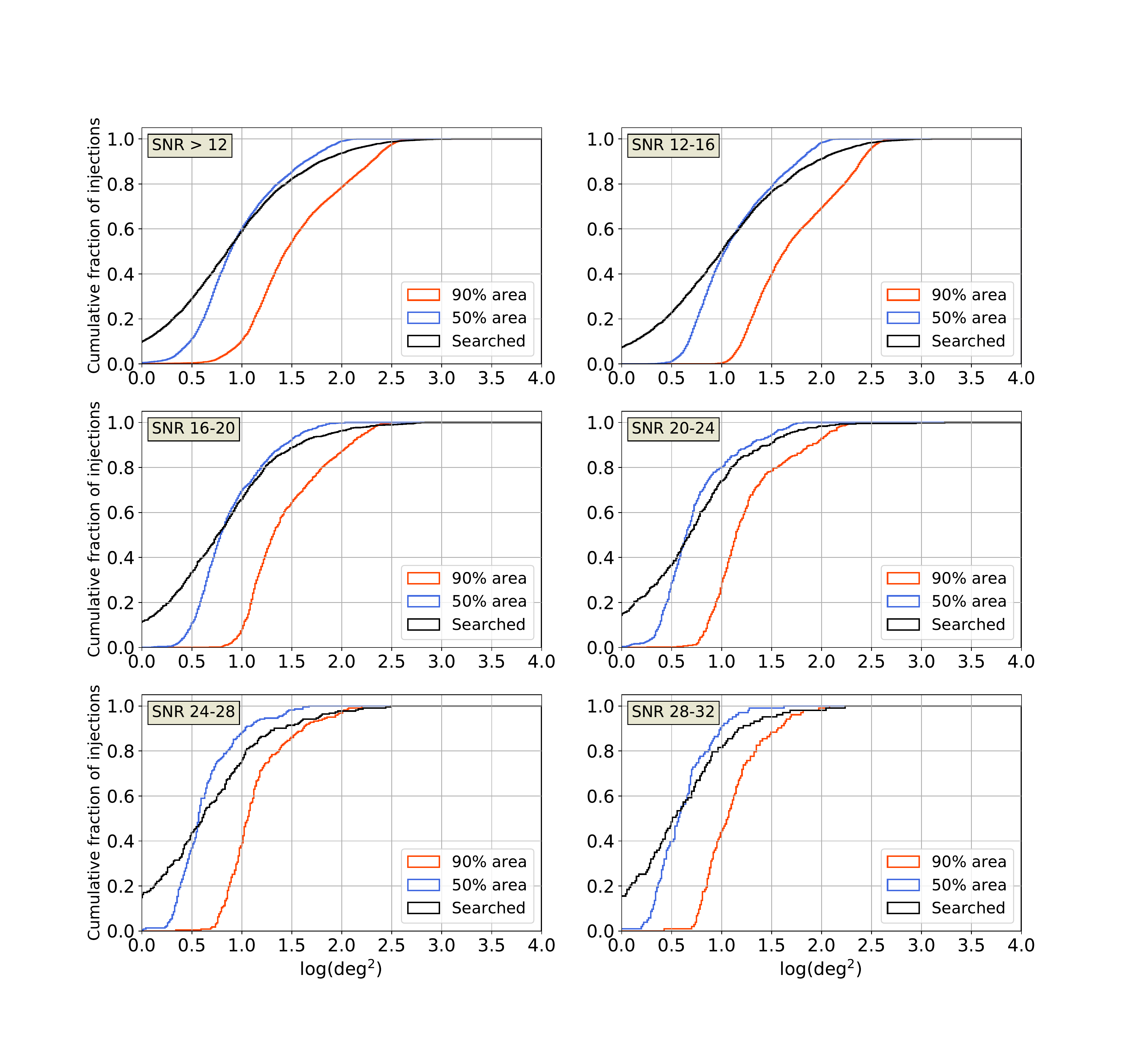}
    \caption{\label{onlyarea_design} Cumulative distribution of the searched area, 90\% confidence area and 50\% confidence area of design sensitivity. Left top panel is for all detections; others are results in various SNR bins.  }
\end{figure*}

\begin{figure*}
    \includegraphics[width=1\textwidth]{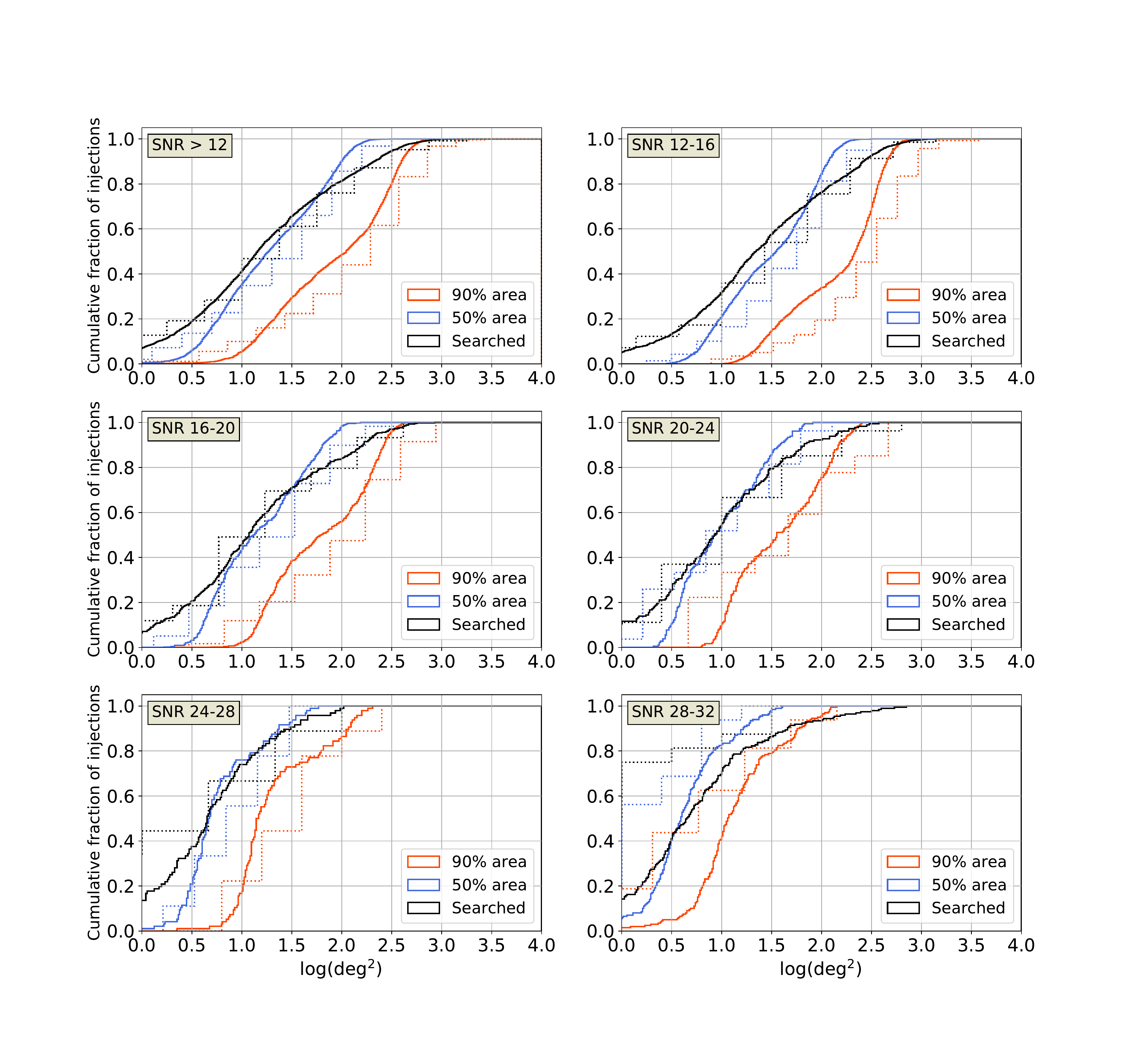}
    \caption{\label{onlyarea_o2} Cumulative distribution of the searched area, 90\% confidence area and 50\% confidence area for the localization of injected GW signals on random Gaussian noise with the smoothed O2 HLV PSD (described in Sec.~\ref{sec3}). We compare our results (solid lines) with \texttt{LALInference} (dotted lines) under the same detector configuration and SNR ranges. The two methods are consistent in most SNR ranges, while the difference in the right bottom panel is distinct due to \texttt{LALInference}'s small sample sizes ($<$10).  }
\end{figure*}

Cumulative distributions of the searched area, 90\% confidence area and 50\% confidence area for different SNR ranges are shown in Fig.~\ref{onlyarea_design} and \ref{onlyarea_o2} for HLV design and O2 sensitivity, respectively. As anticipated, these areas decrease with the increase of SNR, and injections on design sensitivity give much smaller areas. Medians of 90\% confidence area for design sensitivity are $\sim \mathcal{O}(10^1) ~\mathrm{deg}^2$ for SNR $\ge$ 12, while in O2 sensitivity is $\sim \mathcal{O}(10^2) ~\mathrm{deg}^2$. This is consistent with the previous result \texttt{Bayestar}~\cite{Singer2016} where the median for the O2 sensitivity for SNR $\ge$ 12 is about 200~$\mathrm{deg}^2$. In Fig.~\ref{onlyarea_o2}, we compare our results with those of \texttt{LALInference}'s, which are obtained from the data release of Ref.~\cite{Berry2015,Singer2014}. The statistics of both of the methods are generated with the same detector configuration and SNR ranges. It shows the area cumulative distributions of our algorithm are generally comparable to that of \texttt{LALInference}. In the SNR range of 28 to 32, the statistic errors of \texttt{LALInference}'s results become large due to their small sizes of samples ($<$10), so an observable difference between the two methods comes up. The mismatch is supposed to shrink with the increase in the sample size. 

\subsection{Self-consistency\label{sec5b}}
The p-p plot, widely used in GW astronomy~\cite{Sidery_2014,Singer2016}, is used to check the self-consistency of the confidence area statistically. For a given confidence level, we calculate the percentage of random injections with their true sky direction falling within the confidence area and plot the number against the confidence level. We expect these two numbers to be roughly the same, that is, we expect a diagonal line in the p-p plot for perfectly estimated confidence levels.

\begin{figure*}
    \includegraphics[width=1\textwidth]{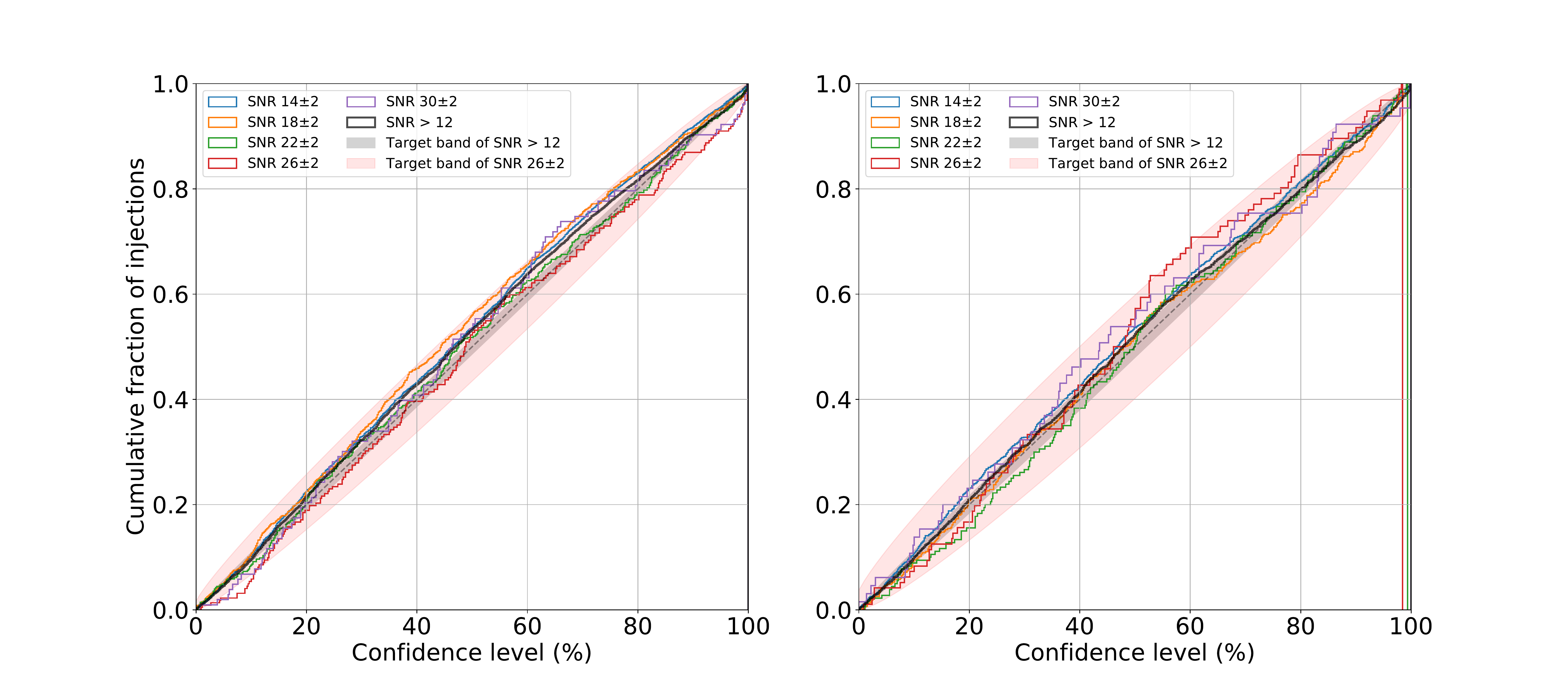}
    \caption{\label{ppplot} P-P plots of our localization algorithm in different SNR ranges. Left panel shows design sensitivity and right panel shows O2 sensitivity explained in Sec.~\ref{sec3}. The gray and red regions around the diagonal are the target 95\% confidence bands for different SNR ranges derived from a binomial distribution.}
\end{figure*}

Our p-p plots are presented in Fig.~\ref{ppplot}. Among all SNR ranges, p-p plots of high SNR cases (e.g. SNR$\sim$26) are more likely to deviate from the diagonal line, as indicated in Fig.~\ref{ppplot}. It \textcolor{black}{is} caused by insufficient high-SNR injection events. Since all parameters are sampled randomly, the high-SNR events are from a shorter distance and therefore constitute a smaller volume of the GW sources. Evidently, the errors become larger for both empirical relation fitting (Eqs.~\ref{emp1} and \ref{emp2}) and the percentage calculation in the p-p plot. For the O2 sensitivity, the total number of high SNR events is less than that in design sensitivity, therefore, the deviation is more distinct.

Following Ref.~\cite{Singer2016}, we plot the 95\% target confidence bands derived from a binomial distribution~\footnote{\url{https://lscsoft.docs.ligo.org/ligo.skymap/plot/pp.html}} for network SNR$\sim$26 and $>$12. Falling in the target band means the error of the p-p plot is acceptable under 95\% confidence.
It shows although the p-p plots of SNR$\sim$26 deviate from the diagonal line, they are still in the target confidence bands which is wide due to the small volume of samples. The overall p-p plot (SNR$>$12) of design sensitivity is slightly out of the target confidence band. We credit this to our way of dealing with the correlation of $A_{ij}$. In Sec.~\ref{sec3}, we assume $A_{22}$ has half chance to be $A_{11}$ and another half to be $-A_{11}$, which is resulted from $|\cos \iota| \rightarrow 1$. When we select high SNR events, sources with $|\cos \iota| \rightarrow 1$ are more likely to be selected. However, if the noise level is very low (e.g. GW detectors achieve the design sensitivity), it does not need large $|\cos \iota|$ to produce high SNR, so that we would select many GW events with small $|\cos \iota|$ and the $A_{22}=\pm A_{11}$ approximation breaks down. This explains why the overall p-p plot is in the target confidence band for O2 sensitivity but slightly deviates for design sensitivity. A simple way to deal with this problem is to give up the correlation assumption and treat all $A_{ij}$s as independent parameters. We derive results from this alternative prior choice in Appendix~\ref{apdxb} which may be useful to the future detectors with better sensitivity. Considering the current detector sensitivity and the performance of our localization method (Fig.~\ref{onlyarea_design},~\ref{onlyarea_o2} and \ref{ppplot}), we still keep the $A_{22}=\pm A_{11}$ approximation.

We should also note that the p-p lines are within the target confidence band and close to the diagonal line at the 90\% confidence level, which means the algorithm is self-consistent for the 90\% confidence area.

\subsection{Real Event Test: GW170817}
GW170817 is the most accurately localized GW event to date. The exact sky direction is obtained via multimessenger observation -- the source is located in NGC 4993 with approximately $(\alpha,\delta) = (197.45^{\circ}, -23.38^{\circ})$~\cite{TheLIGOScientific:2017qsa,Goldstein:2017mmi,Monitor:2017mdv,Savchenko:2017ffs,Abbott2017a}. 

\begin{figure*}
    \includegraphics[width=0.9\textwidth]{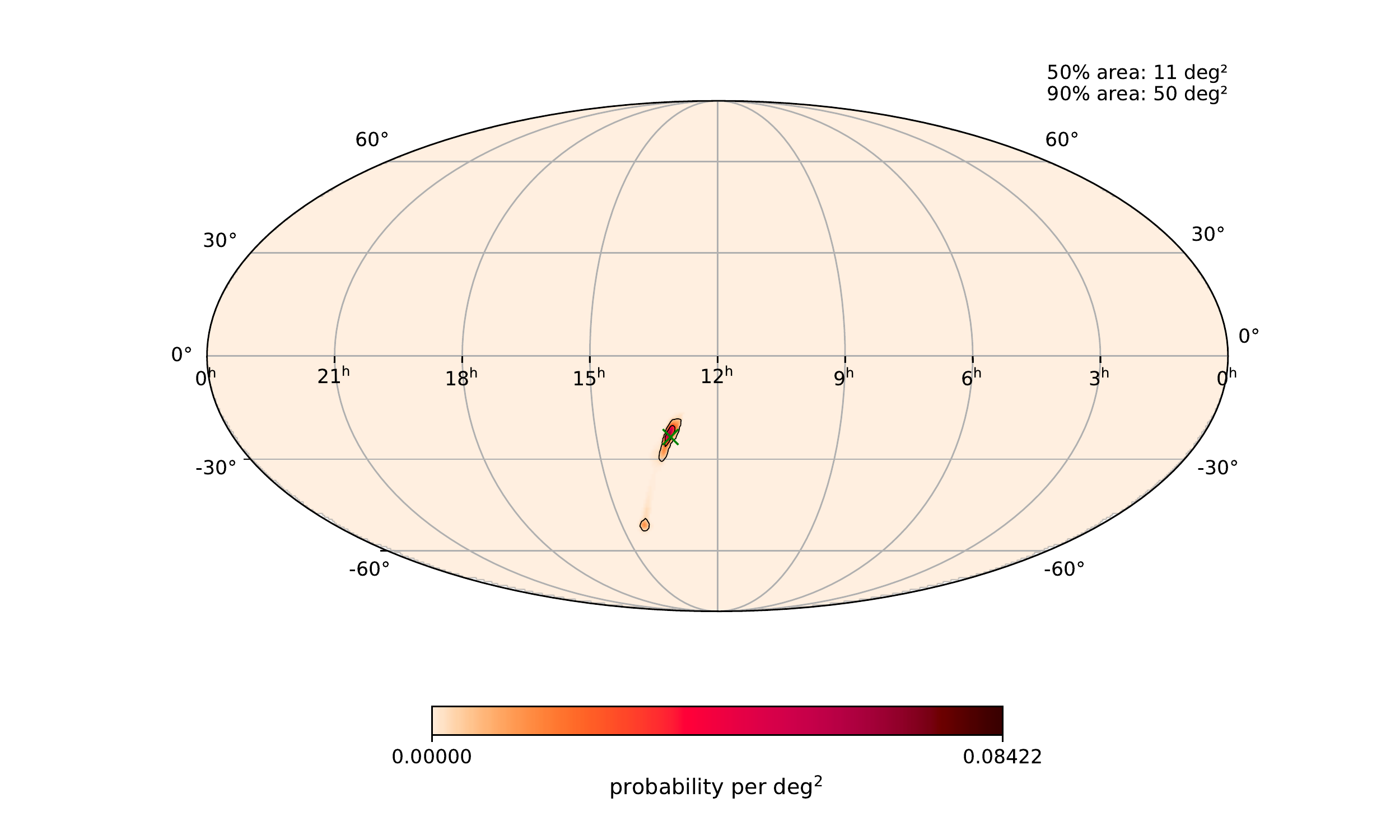}
    \caption{\label{170817} Localization skymap of GW170817 using the SPIIR SNR time-series output. Black line represents 90\% and 50\% confidence contour and a green cross denotes the sky direction taken from the optical observations of the BNS event~\cite{Abbott2017a}. The 90\% credible area is 50$~\mathrm{deg}^2$ and 50\% confidence area is 11$~\mathrm{deg}^2$. Note the true sky direction is within the 50\% confidence area.}
\end{figure*}

\textcolor{black}{We use the public \lwen{cleaned} data of GW170817 from Gravitational Wave Open Science Center~\cite{GWOSC}, and use the SPIIR pipeline~\cite{spiir_paper} to estimate the PSD and 
\lwen{generate the} SNR timeseries. The SPIIR pipeline is also public on Gitlab~\footnote{\url{https://git.ligo.org/lscsoft/spiir}}. We use the PSD near the coalescence time of GW170817 \lwen{and injections described previously} to re-fit the $\mu,\sigma \sim$ SNR relation}: 
\begin{equation}
    \label{bnsfitting}
    \begin{aligned}
        \mu &= 0.0004584~\mathrm{SNR} - 0.0007338,\\
        \sigma &= 0.0002892~\mathrm{SNR} - 0.0004015.
    \end{aligned}
\end{equation}
The horizon distances of H1, L1 and V1 are 213~Mpc, 142~Mpc, 60~Mpc, respectively. We employ the SNR of 3 detectors from GPS time 1187008882.42~s to 1187008882.44~s, and generate the skymap of GW170817, as shown in Fig.~\ref{170817}. The 90\% and 50\% confidence areas are 50 and 11~$\mathrm{deg}^2$, respectively. The optical sky direction (green cross in the plot) is included in both the 90\% and 50\% confidence contour. It shows our algorithm efficiently localized GW170817 with a precision comparable to the rapid localization in real-time detection~\cite{TheLIGOScientific:2017qsa}, where the source is localized to a region of 31~$\mathrm{deg}^2$. 

\subsection{Run Time}
\textcolor{black}{
\lwen{For a given set of 
} SNR timeseries of 3 detectors (\lwen{SNR timeseries have the length of 20 ms and are sampled at 4 kHz} 
), it takes about 2.4 seconds to generate a posterior probability \lwen{distribution} skymap with one thread on Intel(R) Xeon(R) CPU E5-2695 (2.40GHz). \lwen{In comparison}, Bayestar uses $\sim \mathcal{O}(10^2)$ seconds with one \lwen{thread on Intel Xeon E5-2630 v3 CPU}~\cite{Singer2016}. 
\lwen{
} The run time can be further reduced by multi-threads calculation or some other computational techniques.
}

\section{\label{sec6} Conclusions and future work}
We have presented a semianalytical approach (Eq.~\ref{result}) for GW source localization. Compared with the localization algorithm in Ref.~\cite{Singer2016}, our algorithm requires fewer numerical integrations, thus it is expected to reduce the computational cost and latencies in online real-time source localization. In our test with simulated data as well as a case study with GW170817, the algorithm shows fair self-consistency and accuracy. 

\textcolor{black}{Note we assume extrinsic and intrinsic parameters are separable in Eq.~\ref{eq:rearranged} and this is not valid for \lwen{binaries with non-aligned spins}. 
\lwen{We do not expect this to have large impacts on BNS events as most of them are expected to have low spins~\cite{TheLIGOScientific:2017qsa, Abbott2020b,Abbott2019}.
}
}

All required inputs for this algorithm, (the matched filtering SNR time series, event time to calculate the antenna beam pattern functions for the response matrix $\mathbf{G}_\sigma$) in this approach are readily available in online search pipelines SPIIR and are expected to be available for other online search pipelines. It can be used to calculate the posterior probability of source direction following immediately after a GW event is triggered before the event trigger is submitted to the database. It can serve as a complement to the rapid position reconstruction given by \texttt{Bayestar} for GW events submitted to the database.

\begin{acknowledgments}
\textcolor{black}{
We would like to thank Takuya Tsutsui, Kipp Cannon, and Leo Singer for their valuable suggestions. \lwen{We thank} Leo Singer \lwen{for providing the} 
\texttt{LALInference} data for Fig.~\ref{onlyarea_o2}.} LW, QC, and MC acknowledge the funding of Australian Research Council (ARC) Centre of Excellence for Gravitational Wave Discovery OzGrav under grant CE170100004. 

This research was undertaken with the assistance of computational resources from the Pople high-performance computing cluster of the Faculty of Science at the University of Western Australia and the OzStar computer cluster at Swinburne University of Technology.  The OzSTAR program receives funding in part from the Astronomy National Collaborative Research Infrastructure Strategy (NCRIS) allocation provided by the Australian Government.

\end{acknowledgments}

\appendix

\section{Derivation of the analytical marginalization\label{apdx}}
In this appendix we will derive the analytical integral in the posterior probability Eq.~\ref{result}, i.e.,
\begin{equation}
    \begin{aligned}
    I = \int & d^{4} \mathbf{A} \exp \left\{\sum_{x=\{c, s\}}-\frac{1}{2} \mathbf{A}_{\mathbf{x}}^{\mathrm{T}} \mathbf{M} \mathbf{A}_{\mathbf{x}}+\mathbf{J}_{\mathbf{x}}^{\mathrm{T}} \mathbf{A}_{\mathbf{x}}\right\} \\
    & \times p(A_{11}) p(A_{22}|A_{11}) p(A_{21}) p(A_{12}|A_{21})\\
    \end{aligned}
\end{equation}
Here $\mathbf{A_{c}} = (A_{11}, A_{21})^{\mathrm{T}}$, $\mathbf{A_{s}} = (A_{12}, A_{22})^{\mathrm{T}}$, $\mathbf{A}  = (\mathbf{A_{c}},\mathbf{A_{s}})$. For simplicity, define
\begin{equation}
    f_x(p,q) = \exp \left\{-\frac{1}{2} \mathbf{v}^{\mathrm{T}} \mathbf{M} \mathbf{v}+\mathbf{J}_{\mathbf{x}}^{\mathrm{T}} \mathbf{v}\right\} ,
\end{equation}
where $\mathbf{v} = (p,q)^T$, $x=\{c,s\}$. The likelihood Eq.~\ref{LR_re} can be written as $f_c(A_{11}, A_{21})f_s(A_{12},A_{22})$. Note that $f_x(p,q) \neq f_x(q,p)$. The integral $I$ can be rewritten as
\begin{widetext}
\begin{equation}
    \label{marginalizedpost}
    \begin{aligned}
        I& =  \int d^{4} \mathbf{A} f_c(A_{11}, A_{21})f_s(A_{12},A_{22}) \times p(A_{11}) p(A_{22}|A_{11}) p(A_{21}) p(A_{12}|A_{21})\\
        & =  \int d^{4} \mathbf{A} f_c(A_{11}, A_{21})f_s(A_{12},A_{22}) \times  p(A_{11})p(A_{21})  [\delta(A_{22} - A_{11}) + \delta(A_{22}+A_{11})]~ [\delta(A_{12} - A_{21}) + \delta(A_{12}+A_{21})] \\
        & = \int d^{2} \mathbf{A_{c}} p(A_{11})p(A_{21}) \times [ f_c(A_{11}, A_{21})f_s(A_{21}, A_{11})+ f_c(A_{11}, A_{21})f_s(-A_{21}, A_{11}) \\
        & ~~~~~~~~~~~~~~~~~~~~~~~~~~~~~~~~~~~~~~~~+f_c(A_{11}, A_{21})f_s(A_{21}, -A_{11})+f_c(A_{11}, A_{21})f_s(-A_{21}, -A_{11}) ]\\
        & = I_1+I_2+I_3+I_4,
    \end{aligned}
\end{equation}
\end{widetext}
where $ d^{2} \mathbf{A_{c}}=dA_{11}dA_{21}$, $I_{1,2,3,4}$ correspond to the four terms in the 3rd-4th lines. For example,
\begin{equation}
    \begin{aligned}
        I_1 &= \int d^{2} \mathbf{A_{c}} p(A_{11})p(A_{21})f_c(A_{11}, A_{21})f_s(A_{21}, A_{11})\\
        &= \int d^{2} \mathbf{A_{c}} p(A_{11})p(A_{21})\exp \left\{-\frac{1}{2} \mathbf{A_{c}}^{\mathrm{T}} (\mathbf{M + \overline{M})} \mathbf{A_{c}}+\mathbf{J}_{1}^{\mathrm{T}} \mathbf{A_c}\right\},
    \end{aligned}
\end{equation}
where
\begin{equation}
    \mathbf{M} = \mathbf{G^T}_\sigma \mathbf{G}_\sigma =  
    \begin{pmatrix}
        M_{11} & M_{12}\\
        M_{21} & M_{22}
    \end{pmatrix}, ~~
    \mathbf{J_x} = 
    \begin{pmatrix}
        \mathrm{J_{x1} }\\
        \mathrm{J_{x2}}
    \end{pmatrix}.
\end{equation}
are defined in Eq.~\ref{defm} and Eq.~\ref{defj} and
\begin{equation}
    \mathbf{\overline{M}} = 
    \begin{pmatrix}
        M_{22} & M_{12}\\
        M_{21} & M_{11}
    \end{pmatrix}, ~~
    \mathbf{J_1} = 
    \begin{pmatrix}
        \mathrm{J_{c1} + J_{s2}}\\
        \mathrm{J_{c2} + J_{s1}}
    \end{pmatrix}.
\end{equation}
Since $p(A_{11})p(A_{21})$ is the product and summation of several Gaussian functions, the integrand of $I_1$ finally is still a quadratic form on the exponent, thus Eq.~\ref{wiki} can be applied. We finally get
\begin{equation}
    \label{I1_ana}
    I_1 =  \sqrt{\frac{4\pi^2}{\det (\mathbf{ M + \overline{M} )}}} \sum_{k=1}^4 \exp \left\{\frac{1}{2} \mathbf{J_1^{(k)\mathrm{T}}} \mathbf{M}^{\prime-1}\mathbf{J_1^{(k)}} \right\} ,
\end{equation}
where
\begin{equation}
    \mathbf{M'} = \mathbf{ M + \overline{M}} + \frac{1}{\sigma^2} \mathbf{I}, ~~
    \mathbf{J_1^{(k)}} =  \mathbf{J_1} +  \boldsymbol{\alpha^{(k)}}.
\end{equation}
and 
\begin{equation}
    \begin{aligned}
    & \boldsymbol{\alpha^{(1)}} = 
    \begin{pmatrix}
        \mu/\sigma^2 \\
        \mu/\sigma^2 
    \end{pmatrix},
    \boldsymbol{\alpha^{(2)}} = 
    \begin{pmatrix}
        \mu/\sigma^2 \\
        -\mu/\sigma^2 
    \end{pmatrix}, \\
    & \boldsymbol{\alpha^{(3)}} = 
    \begin{pmatrix}
        -\mu/\sigma^2 \\
        \mu/\sigma^2 
    \end{pmatrix},
    \boldsymbol{\alpha^{(4)}} = 
    \begin{pmatrix}
        -\mu/\sigma^2 \\
        -\mu/\sigma^2 
    \end{pmatrix} .
    \end{aligned}
\end{equation}
where $\mu,~\sigma$ are parameters in the bimodal prior function. Eq.~\ref{I1_ana} gives the analytical result of $I_1$. We can solve the other 3 terms with the same method:
\begin{equation}
    \begin{aligned}
        I_2 &= \int d^{2} \mathbf{A_{c}} p(A_{11})p(A_{21})f_c(A_{11}, A_{21})f_s(-A_{21}, A_{11})\\
        &= \int d^{2} \mathbf{A_{c}}  p(A_{11})p(A_{21})\times \\
        &~~~~~~~~~~~  \exp \left\{-\frac{1}{2} \mathbf{A_{c}}^{\mathrm{T}} (\mathbf{M_0 + \overline{M_0})} \mathbf{A_{c}}+\mathbf{J}_{1}^{\mathrm{T}} \mathbf{A_c}\right\} \\
        &= \sqrt{\frac{4\pi^2}{\det (\mathbf{ M_0 + \overline{M_0} )}}} \sum_{k=1}^4 \exp \left\{\frac{1}{2} \mathbf{J_2^{(k)T}} \mathbf{M_0}^{\prime-1} \mathbf{J_2^{(k)}} \right\} ,
    \end{aligned}
\end{equation}
where 
\begin{equation}
    \mathbf{M_0} = 
    \begin{pmatrix}
        M_{11} & 0\\
        0 & M_{22}
    \end{pmatrix}, ~~
    \mathbf{\overline{M_0}} = 
    \begin{pmatrix}
        M_{22} & 0\\
        0 & M_{11}
    \end{pmatrix}, ~~
\end{equation}
\begin{equation}
    \mathbf{J_2} = 
    \begin{pmatrix}
        \mathrm{J_{c1} + J_{s2}}\\
        \mathrm{J_{c2} - J_{s1}}
    \end{pmatrix},
\end{equation}
and
\begin{equation}
    \mathbf{M_0'} = \mathbf{ M_0 + \overline{M_0}} + \frac{1}{\sigma^2} \mathbf{I}, ~~
    \mathbf{J_2^{(k)}} =  \mathbf{J_2} +  \boldsymbol{\alpha^{(k)}}.
\end{equation}
$I_3$ and $I_4$ take similar forms:
\begin{equation}
    \begin{aligned}
        I_3 = \sqrt{\frac{4\pi^2}{\det (\mathbf{ M_0 + \overline{M_0} )}}} \sum_{k=1}^4 \exp \left\{\frac{1}{2} \mathbf{J_3^{(k)\mathrm{T}}} \mathbf{M_0}^{\prime-1} \mathbf{J_3^{(k)}} \right\} ,
    \end{aligned}
\end{equation}
\begin{equation}
    \begin{aligned}
        I_4 = \sqrt{\frac{4\pi^2}{\det (\mathbf{ M + \overline{M} )}}} \sum_{k=1}^4 \exp \left\{\frac{1}{2} \mathbf{J_4^{(k)\mathrm{T}}} \mathbf{M}^{\prime-1}\mathbf{J_4^{(k)}}\right\} ,
    \end{aligned}
\end{equation}
with
\begin{equation}
    \mathbf{J_3} = 
    \begin{pmatrix}
        \mathrm{J_{c1} - J_{s2}}\\
        \mathrm{J_{c2} + J_{s1}}
    \end{pmatrix}, ~~
    \mathbf{J_4} = 
    \begin{pmatrix}
        \mathrm{J_{c1} - J_{s2}}\\
        \mathrm{J_{c2} - J_{s1}}
    \end{pmatrix},
\end{equation}

\begin{equation}
    \mathbf{J_3^{(k)}} =  \mathbf{J_3} +  \boldsymbol{\alpha^{(k)}},~\mathbf{J_4^{(k)}} =  \mathbf{J_4} +  \boldsymbol{\alpha^{(k)}}.
\end{equation}
Since $I_{1,2,3,4}$'s analytical results are given, the posterior probability Eq.~\ref{result} only requires one-fold numerical integral over the coalescence time.

\section{Alternative prior choice for $A_{ij}$ \label{apdxb}}
In Sec.~\ref{sec3}, we point out that diagonal elements of $\mathbf{A}$ are correlated and employ the following approximation (Eq.~\ref{prior_of_A}):
\begin{equation}
    \begin{aligned}
        p(\mathbf{A}) &= p(A_{11}, A_{12}, A_{21}, A_{22}) \\
        &=p(A_{11}, A_{22})p(A_{21}, A_{12}) \\
        &= p(A_{11})p(A_{22}|A_{11}) p(A_{21})p(A_{12}|A_{21}).
    \end{aligned}
\end{equation}
However, this approximation may break down when the noise is very low, as indicated in Sec.~\ref{sec5b}. Here we provide an alternative prior distribution of $\mathbf{A}$ which assumes $A_{ij}$s are independent, i.e.,
\begin{equation}
    \begin{aligned}
        p(\mathbf{A}) &= p(A_{11}, A_{12}, A_{21}, A_{22}) \\
        &=p(A_{11})p(A_{22})p(A_{21})p(A_{12}),
    \end{aligned}
\end{equation}
where $p(A_{ij})$ is given in Eq.~\ref{prior}. The posterior can be obtained by similar Gaussian integral:
\begin{widetext}
\begin{equation}
    \label{alterpost}
    \begin{aligned}
    &p\left(\alpha, \delta \mid \mathbf{d}\right) \\
    & \propto \int d t_{c} \int d^{4} \mathbf{A} \exp \left\{\sum_{x=\{c, s\}}-\frac{1}{2} \mathbf{A}_{\mathbf{x}}^{\mathrm{T}} \mathbf{M} \mathbf{A}_{\mathbf{x}}+\mathbf{J}_{\mathbf{x}}^{\mathrm{T}} \mathbf{A}_{\mathbf{x}}\right\} \prod_{x=\{c, s\}} \left(e^{-\frac{(A_{x1}-\mu)^2}{2\sigma^2}} + e^{-\frac{(A_{x1}+\mu)^2}{2\sigma^2}} \right) \left(e^{-\frac{(A_{x2}-\mu)^2}{2\sigma^2}} + e^{-\frac{(A_{x2}+\mu)^2}{2\sigma^2}} \right)\\
    & \propto \int d t_{c} \prod_{x=\{c, s\}} \int d \mathbf{A}_{\mathbf{x}}\sum_{i=1}^4 \exp \left\{-\frac{1}{2} \mathbf{A}_{\mathbf{x}}^{\mathrm{T}} \mathbf{M}^{\prime\prime} \mathbf{A}_{\mathbf{x}}+\mathbf{J^{(i)\mathrm{T}}}_{\mathbf{x}} \mathbf{A}_{\mathbf{x}}\right\} \\
    &\propto \int d t_{c} \frac{(2 \pi)^{2}}{\operatorname{det} \mathbf{M''} }\prod_{x=\{c, s\}} \sum_{i=1}^4 \exp \left\{ \frac{1}{2} \mathbf{J^{(i)\mathbf{T}}}_{\mathbf{x}} \mathbf{M^{''-1}} \mathbf{J^{(i)}}_{\mathbf{x}}\right\},
    \end{aligned}
\end{equation}
\end{widetext}
where
\begin{equation}
    \mathbf{M}^{\prime\prime}=\mathbf{M}+\frac{1}{\sigma^2} \mathbf{I},
\end{equation}
\begin{equation}
    \mathbf{J^{(i)}_{x}} = \mathbf{J_{x} + \boldsymbol{\alpha^{(i)}}},
\end{equation}
and $\mathbf{M}$ and $\boldsymbol{\alpha^{(i)}}$ have the same definition as in Appendix~\ref{apdx}. Eq.~\ref{alterpost} may be used to replace Eq.~\ref{result} for GW detectors with high sensitivity or low-SNR GW events, in which the $A_{22} = \pm A_{11}$ approximation breaks down.

\bibliography{refs.bib}

\end{document}